\documentclass[preprintnumbers,amsmath,amssymb,floatfix,10pt,prd,twocolumn,
superscriptaddress,nofootinbib]{revtex4}

\usepackage[parfill]{parskip}
\usepackage{graphicx}
\usepackage{amssymb}
\usepackage{epstopdf}
\usepackage{color}
\usepackage{xcolor}
\usepackage{float}
\usepackage{amsmath}
\usepackage{orcidlink}
\usepackage{adjustbox}
\usepackage{hyperref}
\usepackage{bm}
\usepackage{amsthm}

\begin{document}

\title{On-shell Homogeneous Curvature Modes in Quadratic $F(R,\mathcal G)$ Cosmology}

\author{Francisco Tello-Ortiz \orcidlink{0000-0002-7104-5746}}
\email{francisco.tello@ufrontera.cl}
\affiliation{Departamento de Ciencias F\'isicas, Universidad de La Frontera, Casilla 54-D, 4811186 Temuco, Chile.}

\author{B. Mishra}
\email{bivu@hyderabad.bits-pilani.ac.in}
\affiliation{Department of Mathematics,
Birla Institute of Technology and Science-Pilani, Hyderabad Campus, Jawahar Nagar, Kapra Mandal, Medchal District, Telangana 500078, India.}

\author{Y. G\'omez-Leyton\orcidlink{0000-0003-3385-4785}}
\email{ygomez@ucn.cl}
\affiliation{Departamento de F\'isica, Universidad Cat\'olica del Norte, Av. Angamos 0610, Antofagasta, Chile.}

\author{Jean B\'aez Cuevas\orcidlink{0000-0002-3308-3362}}
\email{jean.baez@pucv.cl}
\affiliation{
Instituto de F\'isica, Pontificia Universidad Cat\'olica de Valpara\'iso, Casilla 4950, Valpara\'iso, Chile.}

\begin{abstract}
We study homogeneous cosmological perturbations in the quadratic model
$F(R,\mathcal G)=R+\alpha R^2+\beta\mathcal G^2$. Linearization of
the modified Friedmann equation together with dust conservation yields a
second-order equation for the homogeneous Hubble perturbation, or equivalently
a third-order equation for the matter contrast. We then restrict the
asymptotic analysis to backgrounds that satisfy the zeroth-order equations of
the same quadratic theory. Besides the low-curvature dust branch, which
approaches $H=2/(3t)$, the model admits a nontrivial vacuum de Sitter solution
with $\beta H_{\rm dS}^6=1/96$. Its homogeneous curvature modes are stable
only when $\alpha<-1/(6H_{\rm dS}^2)$. At high curvature the background
equation also admits a vacuum big-rip asymptotic branch beginning with
$H=7/(t_s-t)$; the first corrections are fixed by the quadratic couplings.
The corresponding $\mathcal G^2$-dominated perturbation has one divergent
homogeneous mode. The simultaneous Einstein limit is taken only as a
consistency check and recovers the standard dust behavior.
\end{abstract}

\maketitle

\section{Introduction}

The accelerated expansion of the Universe has motivated a broad class of
modified-gravity models in which the late-time dynamics is generated by
curvature corrections rather than by an explicit dark-energy component
\citep{NojiriOdintsov2005,Cognola2006,NojiriOdintsov2011,NojiriOdintsovOikonomou2017}.  A particularly
useful extension combines the Ricci scalar $R$ with the Gauss--Bonnet
invariant $\mathcal G$, allowing cosmological reconstruction and a rich
set of late-time solutions \citep{Elizalde2010,deLaCruzDombriz2012,Lohakare_2023,Lohakare_2026}.
Quadratic curvature corrections are also independently motivated by the
well-known role of $R^2$ terms in early-Universe cosmology
\citep{Starobinsky1980,DeFeliceTsujikawa2010}.

Future cosmological evolution provides a natural arena in which to test
such theories. Finite-time singularities can be classified according to the
behavior of the scale factor and its derivatives
\citep{NojiriOdintsovTsujikawa2005,NojiriOdintsov2008}, while phantom-like
models may instead lead to little-rip or pseudo-rip evolution without a
finite-time divergence of the scale factor
\citep{Frampton2011,Frampton2012Models,Frampton2012Pseudo}. A representative
higher-curvature model within this setting is
\begin{equation}
F(R,\mathcal G)=R+\alpha R^2+\beta\mathcal G^2,
\label{eq:model}
\end{equation}
which provides a simple arena in which the distinction between matter and
curvature modes can be made explicitly. General $f(R,\mathcal G)$ theories
possess a richer perturbative structure than GR, with additional gravitational
degrees of freedom and nontrivial stability conditions
\citep{DeFeliceGerardSuyama2010,DeFeliceSuyama2010}. Consequently, an
algebraic relation between a homogeneous Hubble perturbation and a matter
contrast cannot be assumed to represent the complete finite-coupling dynamics.

Previous analyses have addressed complementary aspects of this framework. In particular, in Ref.~ \citep{deLaCruzDombriz2012} was studied the stability of cosmological solutions in general $f(R,\mathcal G)$ gravity, deriving the homogeneous perturbation structure and applying it to representative backgrounds. More recently, in Ref.~\citep{Lohakare_2026} was considered the same quadratic choice \eqref{eq:model} from the viewpoint of late-time background evolution and observational constraints. The present work is complementary to both: rather than reconstructing or fitting the background expansion, we keep $\alpha$ and $\beta$ finite in the homogeneous perturbation operator and isolate analytically the additional curvature modes, their asymptotic branches, and the distinct roles of the $R^2$ and $\mathcal G^2$ sectors.\par

The first purpose of this work is to derive the homogeneous linear system of
Eq.~\eqref{eq:model} directly from the cosmological field equation. We show
that the Hubble perturbation obeys a second-order differential equation whose
highest-derivative coefficient is proportional to
$\alpha+16\beta H_0^4$. Together with exact dust conservation, this is
equivalent to a third-order equation for the homogeneous matter contrast. The
result displays explicitly the curvature dynamics that is absent from the
Einstein theory. We also identify the degenerate locus on which the
highest-derivative coefficient vanishes and show that this condition alone does
not yield an algebraic reduction on a time-dependent background.

The second purpose is to determine precisely how the matter-dilution law used
in reduced homogeneous analyses is recovered. In the simultaneous Einstein
limit $\alpha,\beta\to0$, the differential perturbation equation collapses to
an algebraic constraint. The latter agrees with the relation obtained directly
by perturbing the Einstein Friedmann equation. This provides a clean
separation between the Einstein-connected matter mode and the additional
finite-coupling curvature modes.

A central point of the present analysis is that perturbative stability is
meaningful only after the zeroth-order equations have been imposed. We
therefore do not treat arbitrary little-rip, pseudo-rip, or power-law
histories as solutions of the quadratic model. Instead, after deriving the
general homogeneous perturbation operator, we solve the background equation
asymptotically and retain only branches supported by the same
$F(R,\mathcal G)$ dynamics and matter content. This on-shell restriction
distinguishes the analysis below from a purely kinematical evaluation of the
perturbation operator.

\section{Full homogeneous perturbation system}
\label{sec:setup}

We consider the spatially flat FLRW line element
\begin{equation}
ds^2=-dt^2+a^2(t)(dx^2+dy^2+dz^2)
\end{equation}
and introduce homogeneous perturbations according to
\begin{equation}
H(t)=H_0(t)\bigl[1+\delta(t)\bigr],\qquad
\rho_m(t)=\rho_{m0}(t)\bigl[1+\delta_m(t)\bigr].
\label{eq:pertdefs}
\end{equation}
For minimally coupled pressureless matter the conservation equation is exact,
\begin{equation}
\dot\rho_m+3H\rho_m=0,
\label{eq:fullcont}
\end{equation}
and its background and linear parts are
\begin{equation}
\dot\rho_{m0}+3H_0\rho_{m0}=0,
\qquad
\dot\delta_m+3H_0\delta=0.
\label{eq:cont}
\end{equation}
Thus the matter perturbation cannot be evolved independently of the prescribed
background expansion.

For a general $F(R,\mathcal G)$ theory, the flat-FLRW Friedmann equation can be
written as
\begin{equation}
3H^2F_R=\kappa^2\rho_m+
\frac12\left(RF_R+\mathcal G F_{\mathcal G}-F\right)
-3H\dot F_R-12H^3\dot F_{\mathcal G}.
\label{eq:fullFriedmann}
\end{equation}
For
\begin{equation}
F(R,\mathcal G)=R+\alpha R^2+\beta\mathcal G^2,
\label{eq:modelagain}
\end{equation}
one has
\begin{equation}
F_R=1+2\alpha R,\qquad F_{\mathcal G}=2\beta\mathcal G,
\end{equation}
so Eq.~\eqref{eq:fullFriedmann} becomes
\begin{equation}
\begin{split}
3H^2(1+2\alpha R)={}&\kappa^2\rho_m+
\frac{\alpha}{2}R^2+\frac{\beta}{2}\mathcal G^2\\
&-6\alpha H\dot R-24\beta H^3\dot{\mathcal G}.
\end{split}
\label{eq:quadraticFriedmann}
\end{equation}
This equation, rather than an assumed algebraic relation between $\delta$ and
$\delta_m$, is the starting point of the homogeneous perturbation analysis.

The curvature invariants are
\begin{equation}
R=6(2H^2+\dot H),\qquad
\mathcal G=24H^2(H^2+\dot H).
\end{equation}
At first order,
\begin{equation}
\delta R=6\left[(4H_0^2+\dot H_0)\delta+H_0\dot\delta\right],
\label{eq:deltaR}
\end{equation}
and
\begin{equation}
\delta\mathcal G=24H_0^2\left[(4H_0^2+3\dot H_0)\delta
+H_0\dot\delta\right].
\label{eq:deltaG}
\end{equation}
The derivatives entering Eq.~\eqref{eq:quadraticFriedmann} also require
\begin{equation}
\begin{split}
\delta\dot R=6\bigl[&H_0\ddot\delta
+(4H_0^2+2\dot H_0)\dot\delta\\
&+(8H_0\dot H_0+\ddot H_0)\delta\bigr],
\end{split}
\label{eq:deltadotR}
\end{equation}
and
\begin{equation}
\begin{split}
\delta\dot{\mathcal G}=24\bigl[&H_0^3\ddot\delta
+(4H_0^4+6H_0^2\dot H_0)\dot\delta\\
&+(16H_0^3\dot H_0+6H_0\dot H_0^2
+3H_0^2\ddot H_0)\delta\bigr].
\end{split}
\label{eq:deltadotG}
\end{equation}
Subtracting the background equation from the linearized
Eq.~\eqref{eq:quadraticFriedmann} gives the full homogeneous constraint
\begin{equation}
\mathcal A_2\ddot\delta+\mathcal A_1\dot\delta
+\mathcal A_0\delta=\kappa^2\rho_{m0}\delta_m,
\label{eq:fullpert}
\end{equation}
where
\begin{equation}
\mathcal A_2=36H_0^2\left(\alpha+16\beta H_0^4\right),
\label{eq:A2}
\end{equation}
\begin{equation}
\mathcal A_1=36H_0\left[
\alpha(3H_0^2+\dot H_0)
+16\beta H_0^4(3H_0^2+5\dot H_0)
\right],
\label{eq:A1}
\end{equation}
and
\begin{equation}
\begin{split}
\mathcal A_0={}&6H_0^2
+324\alpha H_0^2\dot H_0
+72\alpha H_0\ddot H_0
-36\alpha\dot H_0^2\\
&-2304\beta H_0^8
+12096\beta H_0^6\dot H_0
+3456\beta H_0^5\ddot H_0\\
&+5184\beta H_0^4\dot H_0^2.
\end{split}
\label{eq:A0}
\end{equation}
Equations~\eqref{eq:cont} and \eqref{eq:fullpert} constitute the homogeneous
linear system. In particular, for generic nonzero $\alpha$ and $\beta$ the
Hubble perturbation is dynamical: the quadratic curvature sector cannot in
general be replaced by a single algebraic closure.

\subsection{Matter-only form and curvature modes}

Equation~\eqref{eq:cont} permits the elimination of the Hubble perturbation,
\begin{equation}
\delta=-\frac{\dot\delta_m}{3H_0}.
\label{eq:deltafrommatter}
\end{equation}
Substitution into Eq.~\eqref{eq:fullpert} gives an equivalent third-order
equation for the homogeneous matter contrast,
\begin{equation}
\mathcal A_2\frac{d^2}{dt^2}\left(\frac{\dot\delta_m}{H_0}\right)
+\mathcal A_1\frac{d}{dt}\left(\frac{\dot\delta_m}{H_0}\right)
+\mathcal A_0\frac{\dot\delta_m}{H_0}
+3\kappa^2\rho_{m0}\delta_m=0.
\label{eq:thirdorder}
\end{equation}
The order of Eq.~\eqref{eq:thirdorder} makes explicit that the complete
quadratic theory contains homogeneous curvature dynamics in addition to the
matter-dilution mode. This is the essential distinction between the nonzero
coupling theory and its Einstein limit.

\subsection{Degenerate highest-derivative locus}

The coefficient of the highest derivative vanishes when
\begin{equation}
\alpha+16\beta H_0^4=0.
\label{eq:degenerate}
\end{equation}
This condition alone does not generically produce an algebraic perturbation
constraint. Indeed, on the locus \eqref{eq:degenerate},
\begin{equation}
\mathcal A_1=2304\beta H_0^5\dot H_0.
\label{eq:A1deg}
\end{equation}
For $\beta\neq0$, the first-derivative term therefore survives whenever the
background is time dependent. A genuinely algebraic reduction requires, in
addition, $\dot H_0=0$ (or a trivial coupling limit). For a constant-Hubble
background satisfying Eq.~\eqref{eq:degenerate},
\begin{equation}
\mathcal A_0=6H_0^2\left(1-384\beta H_0^6\right),
\end{equation}
and Eq.~\eqref{eq:fullpert} becomes algebraic. This observation is a statement
about the perturbation operator. An exact de Sitter background sourced only by
nonvanishing conserved dust is not implied: since $\rho_{m0}\propto a^{-3}$,
such a realization requires the dust to be absent, subdominant, or accompanied
by an additional component.

\section{On-shell finite-coupling backgrounds and their homogeneous modes}
\label{sec:finitecoupling}

We now keep the quadratic couplings finite and, crucially, impose the
zeroth-order background equation before discussing perturbations. For
conserved dust,
\begin{equation}
\dot\rho_{m0}+3H_0\rho_{m0}=0,
\qquad
\rho_{m0}=\rho_{m*}a^{-3},
\label{eq:dustbackground}
\end{equation}
while Eq.~\eqref{eq:quadraticFriedmann} is equivalently
\begin{align}
\kappa^2\rho_{m0}={}&3H_0^2
+108\alpha H_0^2\dot H_0
+36\alpha H_0\ddot H_0
-18\alpha\dot H_0^2 \nonumber\\
&-288\beta H_0^8
+1728\beta H_0^6\dot H_0
+576\beta H_0^5\ddot H_0
 \nonumber\\
&+864\beta H_0^4\dot H_0^2 .
\label{eq:bgexpanded}
\end{align}
The asymptotic branches studied below are selected from
Eqs.~\eqref{eq:dustbackground}--\eqref{eq:bgexpanded}, hence the perturbation
operator is evaluated on backgrounds of the same theory.

\subsection{Vacuum de Sitter branch}

Let
\begin{equation}
H_0=H_{\rm dS}>0,\qquad
\dot H_0=\ddot H_0=0,\qquad \rho_{m0}=0 .
\end{equation}
Equation~\eqref{eq:bgexpanded} gives
\begin{equation}
3H_{\rm dS}^2-288\beta H_{\rm dS}^8=0 .
\end{equation}
Besides Minkowski space, the nontrivial de Sitter branch therefore exists
only for
\begin{equation}
{\beta H_{\rm dS}^6=\frac{1}{96}},
\label{eq:dSonshell}
\end{equation}
which in particular requires $\beta>0$.

On this exact background the perturbation coefficients are
\begin{align}
\mathcal A_2^{\rm dS}
&=36H_{\rm dS}^2
\left(\alpha+16\beta H_{\rm dS}^4\right),\\
\mathcal A_1^{\rm dS}
&=108H_{\rm dS}^3
\left(\alpha+16\beta H_{\rm dS}^4\right),\\
\mathcal A_0^{\rm dS}
&=6H_{\rm dS}^2
\left(1-384\beta H_{\rm dS}^6\right).
\end{align}
Away from the degenerate locus, $\delta\propto e^{\lambda t}$ satisfies
\begin{equation}
\lambda^2+3H_{\rm dS}\lambda+\mu_{\rm dS}^2=0,
\qquad
\mu_{\rm dS}^2=
\frac{1-384\beta H_{\rm dS}^6}
{6(\alpha+16\beta H_{\rm dS}^4)} .
\label{eq:muds}
\end{equation}
Using the on-shell condition \eqref{eq:dSonshell},
\begin{equation}
\mu_{\rm dS}^2=
-\frac{1}{2\left(\alpha+1/(6H_{\rm dS}^2)\right)} .
\label{eq:mudsonshell}
\end{equation}
Since $H_{\rm dS}>0$, both homogeneous curvature modes have negative real
part precisely when
\begin{equation}
{\alpha<-\frac{1}{6H_{\rm dS}^2}} .
\label{eq:dSonshellstability}
\end{equation}
The equality is degenerate and must be analyzed separately. Equation
\eqref{eq:dSonshellstability} is an on-shell homogeneous-stability
condition; it should not be interpreted as a complete no-ghost condition
for the scalar, vector, and tensor perturbation sectors of the theory.

\subsection{Low-curvature dust branch}

At late times a dust solution with decreasing curvature can approach the
Einstein matter-dominated behavior. Let
\begin{equation}
H_0(t)\sim\frac{h}{t},\qquad t\rightarrow\infty .
\end{equation}
Then $a\sim t^h$ and conservation gives
$\rho_{m0}\sim\rho_*t^{-3h}$. The Einstein term in
Eq.~\eqref{eq:bgexpanded} is of order $t^{-2}$, the $R^2$ corrections start
at order $t^{-4}$, and the $\mathcal G^2$ corrections at order $t^{-8}$.
Matching the leading dust and Einstein powers therefore requires
\begin{equation}
3h=2,\qquad h=\frac23,
\label{eq:dustbranch}
\end{equation}
with $\kappa^2\rho_*=4/3$ in the normalization $a\sim t^{2/3}$.
Consequently
\begin{equation}
H_0(t)=\frac{2}{3t}+O(t^{-3})
\end{equation}
is a genuine low-curvature asymptotic branch of the quadratic theory; the
higher-curvature terms determine subleading corrections rather than changing
the leading dust exponent.

For fixed $\alpha\neq0$, the highest-derivative part of the perturbation
operator is controlled asymptotically by the $R^2$ sector, while the
$\mathcal G^2$ terms are further suppressed. Equation~\eqref{eq:fullpert}
therefore reduces at leading curvature-mode order to
\begin{equation}
\ddot\delta+3H_0\dot\delta+\frac{1}{6\alpha}\delta\simeq0 .
\label{eq:dustcurvaturemode}
\end{equation}
For $\alpha>0$ the additional homogeneous mode is asymptotically
oscillatory, with WKB envelope
\begin{equation}
|\delta|\propto a^{-3/2}\propto t^{-1},
\end{equation}
whereas $\alpha<0$ gives a tachyonic branch
\begin{equation}
\delta\sim
t^{-1}\exp\!\left(\frac{t}{\sqrt{6|\alpha|}}\right)
\end{equation}
up to subleading corrections. Thus the low-curvature dust branch separates
the two signs of the $R^2$ coupling already at homogeneous level.

\subsection{High-curvature vacuum big-rip branch}

The background equation itself selects a nontrivial high-curvature
finite-time branch. Write
\begin{equation}
\tau=t_s-t,\qquad
H_0(\tau)\sim\frac{\gamma}{\tau},
\qquad \tau\rightarrow0^+ .
\end{equation}
For $\beta\neq0$ the leading contribution to
Eq.~\eqref{eq:bgexpanded} is
\begin{equation}
\kappa^2\rho_{m0}
\sim
-288\beta\gamma^6(\gamma-7)(\gamma+1)\tau^{-8}.
\label{eq:BRbackgroundleading}
\end{equation}
Conserved dust behaves as $\rho_{m0}\propto\tau^{3\gamma}$ and is
negligible at the endpoint. A positive expanding vacuum branch therefore
requires
\begin{equation}
{\gamma=7}.
\label{eq:gamma7}
\end{equation}
The cancellation of the $\tau^{-8}$ term is not by itself sufficient:
subleading orders must also satisfy the background equation. They do. An
asymptotic expansion gives
\begin{equation}
H_0(\tau)=\frac7{\tau}
+\frac{3\alpha}{87808\,\beta}\tau^3
+O(\tau^5).
\label{eq:BRseries}
\end{equation}
Substitution into the background equation shows that the expansion can be
continued consistently order by order. Thus $H_0\sim7/\tau$ is not merely
a kinematical ansatz: it is the leading term of an asymptotic vacuum
solution of the same quadratic model.

The leading perturbation coefficients on a general
$H_0=\gamma/\tau$ high-curvature branch are
\begin{align}
\mathcal A_2&\sim576\beta\gamma^6\tau^{-6},\\
\mathcal A_1&\sim576\beta\gamma^6(3\gamma+5)\tau^{-7},\\
\mathcal A_0&\sim-576\beta\gamma^6
(4\gamma^2-21\gamma-21)\tau^{-8}.
\end{align}
For $\delta\sim\tau^s$, with $d/dt=-d/d\tau$, the correct indicial equation
is
\begin{equation}
s^2-(3\gamma+6)s
-4\gamma^2+21\gamma+21=0 .
\label{eq:BRbetaindicial}
\end{equation}
Evaluating it on the background-selected value $\gamma=7$ gives
\begin{equation}
s^2-27s-28=0,
\end{equation}
and hence
\begin{equation}
s_+=28,\qquad s_-=-1.
\label{eq:BRroots7}
\end{equation}
Because $s_-<0<s_+$, the on-shell big-rip branch contains one divergent
homogeneous curvature mode as $\tau\to0^+$. It is therefore unstable within
the homogeneous sector.

\subsection{On-shell finite-coupling picture}

The background equation and the perturbation equation together lead to a
more restrictive picture than a classification based on arbitrary
kinematical histories. The quadratic model possesses an exact vacuum de
Sitter branch fixed by $\beta H_{\rm dS}^6=1/96$, a low-curvature dust
branch approaching $H_0=2/(3t)$, and a high-curvature vacuum big-rip
asymptotic branch beginning with $H_0=7/(t_s-t)$. Their homogeneous
stability properties are qualitatively different. The de Sitter branch is
stable only in the parameter range \eqref{eq:dSonshellstability}; the
low-curvature dust branch has an oscillatory additional mode for
$\alpha>0$ and a tachyonic one for $\alpha<0$; and the on-shell big-rip
branch possesses one divergent $\mathcal G^2$-dominated mode.

Little-rip and generic prescribed power-law histories are not included as
stability examples because, with conserved dust alone, they do not satisfy
Eq.~\eqref{eq:bgexpanded} asymptotically for fixed nonzero quadratic
couplings. This restriction ensures that every stability statement made in
this section refers to a background of the same theory being perturbed.

\section{Einstein limit as a consistency check}
\label{sec:GRcheck}

The finite-coupling analysis above is the main result. We now take the
simultaneous Einstein limit only as a consistency check. Setting
$\alpha,\beta\to0$ in Eq.~\eqref{eq:fullpert} gives
\begin{equation}
\mathcal A_2\to0,\qquad \mathcal A_1\to0,\qquad
\mathcal A_0\to6H_0^2,
\end{equation}
so that
\begin{equation}
6H_0^2\delta=\kappa^2\rho_{m0}\delta_m.
\label{eq:GRconstraintcheck}
\end{equation}
For a dust background in general relativity,
\begin{equation}
3H_0^2=\kappa^2\rho_{m0},\qquad
-2\dot H_0=\kappa^2\rho_{m0},
\end{equation}
and therefore $\delta=\delta_m/2$. Together with dust conservation this
implies
\begin{equation}
\dot\delta_m+\frac32H_0\delta_m=0,
\qquad \delta_m\propto a^{-3/2}.
\label{eq:GRmattercheck}
\end{equation}
Thus the algebraic closure is recovered as the Einstein degeneration of the
finite-coupling differential system; it is not imposed on the quadratic
theory. No separate GR classification is required for the finite-coupling
results of this work.

\section{Conclusions}

We have derived the homogeneous perturbation system of
$F(R,\mathcal G)=R+\alpha R^2+\beta\mathcal G^2$ while keeping both
quadratic couplings finite. The modified Friedmann equation yields a
second-order equation for the Hubble perturbation, and dust conservation
converts the system into a third-order equation for the matter contrast.
The additional solutions are homogeneous curvature modes that disappear
when the differential operator degenerates in the Einstein limit.

A central requirement of the analysis has been to impose the zeroth-order
field equation before assigning physical meaning to the perturbation
spectrum. This selects three useful asymptotic branches. The nontrivial
vacuum de Sitter solution obeys $\beta H_{\rm dS}^6=1/96$ and is
homogeneously stable only for
$\alpha<-1/(6H_{\rm dS}^2)$. At low curvature the conserved-dust branch
approaches $H_0=2/(3t)$; its additional $R^2$ mode is oscillatory and
Hubble-damped for $\alpha>0$, while $\alpha<0$ produces a tachyonic
homogeneous branch. At high curvature the vacuum background equation
selects a big-rip asymptotic solution with
$H_0=7/(t_s-t)+O[(t_s-t)^3]$. Its first subleading coefficients are fixed
by $\alpha$ and $\beta$, and the corresponding $\mathcal G^2$-dominated
indicial equation has one negative exponent, establishing homogeneous
instability of that branch.

The general-relativistic calculation is retained only as a consistency
check. Taking $\alpha,\beta\to0$ removes the higher-derivative part of the
homogeneous operator and recovers the standard matter-dominated behavior.
Conversely, arbitrary little-rip and generic prescribed power-law histories
are not used as stability examples because they do not satisfy the
dust-only background equation of the fixed quadratic model. This on-shell
restriction separates genuine stability statements from purely
kinematical tests of the perturbation operator.

Finally, homogeneous stability is not a complete viability criterion for a
higher-curvature theory. The conditions derived here concern the spatially
homogeneous sector and do not by themselves establish the absence of
ghosts or gradient instabilities in the full scalar, vector, and tensor
perturbation spectrum.

\section*{ACKNOWLEDGMENTS}

Y. Gómez-Leyton acknowledges ANID, Subvención a la Academia, convocatoria año 2025, Grant No.~85250186.

\bibliographystyle{apsrev4-2}
\bibliography{references}

\end{document}